\documentclass[%
 reprint,
 amsmath,amssymb,
 aps, prx,
 longbibliography,
 superscriptaddress,
 altaffilsymbol
]{revtex4-1}
\usepackage[left=2cm, right=2cm, top=3cm, bottom=3cm]{geometry}
\usepackage{amsmath,amssymb}
\usepackage{amsopn}
\usepackage{amsfonts}
\usepackage{natbib,graphicx,url,times,layouts,xcolor}
\usepackage{bm}
\usepackage{dsfont}
\usepackage{bbold}
\usepackage{txfonts} % For PRL fonts
\usepackage{cases} % For numbering in cases format
\usepackage{placeins} %FloatBarrier
\usepackage{textcomp} %mu and other symbols
\usepackage{lineno}
\definecolor{blueviolet}{rgb}{0.2, 0.2, 0.6}
\usepackage[pdftex, % FOR PDFLATEX ONLY
bookmarks=false, % Bookmark bar
colorlinks=true,
allcolors=blueviolet,
pdfstartview={FitH}, % FitBH
]{hyperref}

\newcommand{\bra}[1]{{\left\langle{#1}\right\vert}}
\newcommand{\ket}[1]{{\left\vert{#1}\right\rangle}}

\begin{document}
\title{Radiative cooling of a spin ensemble}

\author{B.~Albanese}
\affiliation{Quantronics group, SPEC, CEA, CNRS, Universit\'e Paris-Saclay, CEA Saclay 91191 Gif-sur-Yvette Cedex, France}

\author{S.~Probst}
\affiliation{Quantronics group, SPEC, CEA, CNRS, Universit\'e Paris-Saclay, CEA Saclay 91191 Gif-sur-Yvette Cedex, France}

\author{V.~Ranjan}
\affiliation{Quantronics group, SPEC, CEA, CNRS, Universit\'e Paris-Saclay, CEA Saclay 91191 Gif-sur-Yvette Cedex, France}

\author{C.~Zollitsch}
\affiliation{London Centre for Nanotechnology, University College London, London WC1H 0AH, United Kingdom}

\author{M.~Pechal}
\affiliation{Department of Physics, ETH Zurich, CH-8093 Zurich, Switzerland}

\author{A.~Wallraff}
\affiliation{Department of Physics, ETH Zurich, CH-8093 Zurich, Switzerland}

\author{J.J.L.~Morton}
\affiliation{London Centre for Nanotechnology, University College London, London WC1H 0AH, United Kingdom}

\author{D.~Vion}
\affiliation{Quantronics group, SPEC, CEA, CNRS, Universit\'e Paris-Saclay, CEA Saclay 91191 Gif-sur-Yvette Cedex, France}

\author{D.~Esteve}
\affiliation{Quantronics group, SPEC, CEA, CNRS, Universit\'e Paris-Saclay, CEA Saclay 91191 Gif-sur-Yvette Cedex, France}

\author{E.~Flurin}
\affiliation{Quantronics group, SPEC, CEA, CNRS, Universit\'e Paris-Saclay, CEA Saclay 91191 Gif-sur-Yvette Cedex, France}

\author{P.~Bertet*}
\affiliation{Quantronics group, SPEC, CEA, CNRS, Universit\'e Paris-Saclay, CEA Saclay 91191 Gif-sur-Yvette Cedex, France}

\newcommand{\Tphon}{T_\text{phon}}
\newcommand{\Tphot}{T_\text{phot}}
\newcommand{\Tphotc}{T_\text{phot}^\text{cold}}
\newcommand{\Tphoth}{T_\text{phot}^\text{hot}}
\newcommand{\Ti}{T_\text{int}}
\newcommand{\Tih}{T_\text{int}^\text{hot}}
\newcommand{\Tic}{T_\text{int}^\text{cold}}

\newcommand{\Tc}{T_\text{cold}}
\newcommand{\Ts}{T_\text{spin}}
\newcommand{\Tsc}{T_\text{spin}^\text{cold}}
\newcommand{\Tsh}{T_\text{spin}^\text{hot}}

\newcommand{\ki}{\kappa_\text{int}}
\newcommand{\ke}{\kappa_\text{ext}}

\newcommand{\Gphot}{\Gamma_\text{phot}}
\newcommand{\Gphon}{\Gamma_\text{phon}}
\newcommand{\Gphotc}{\Gamma_\text{phot}^\text{cold}}
\newcommand{\Gphoth}{\Gamma_\text{phot}^\text{hot}}
\newcommand{\Gi}{\Gamma_\text{1}}

\begin{abstract}
    
Physical systems reach thermal equilibrium through energy exchange with their environment, and for spins in solids the relevant environment is almost always the host lattice in which they sit. However, recent studies motivated by observations from Purcell showed how coupling to a cavity can become the dominant form of relaxation for spins, given suitably strong spin-cavity coupling. In this regime, the cavity electromagnetic field takes over from the lattice as the dominant environment, inviting the prospect of controlling the spin temperature independently from that of the lattice, by engineering a suitable cavity field. Here, we report on precisely such control over spin temperature, illustrating a novel and universal method of electron spin hyperpolarisation. By switching the cavity input between loads at different temperatures we can control the electron spin polarisation, cooling it below the lattice temperature. Our demonstration uses donor spins in silicon coupled to a superconducting micro-resonator and we observe an increase of spin polarisation of over a factor of two. This approach provides general route to signal enhancement in electron spin resonance, or indeed nuclear magnetic resonance through dynamical nuclear spin polarisation (DNP).
\end{abstract}

\maketitle

When a physical system is coupled to several reservoirs at different temperatures, it equilibrates at an intermediate temperature which depends on the strength with which it is coupled to each bath. An electron spin in a solid is coupled to two different reservoirs: phonons in its host lattice, and microwave photons in its electromagnetic environment. The strength of this coupling is characterized by the rate at which the spin, of Larmor frequency $\omega_\text{spin}$, relaxes by spontaneously emitting a quantum of energy $\hbar \omega_\text{spin}$ into each bath. In usual magnetic resonance experiments\,\cite{schweiger_principles_2001}, the spin-lattice relaxation rate $\Gamma_\text{phon}$ is many orders of magnitude larger than the radiative relaxation rate $\Gamma_\text{phot}$, so that the spin temperature $T_\text{spin}$ is determined by the sample temperature $T_\text{phon}$ regardless of the temperature of the microwave photons $T_\text{phot}$. 

\begin{figure}[tbph]
  \includegraphics[width=8.2cm]{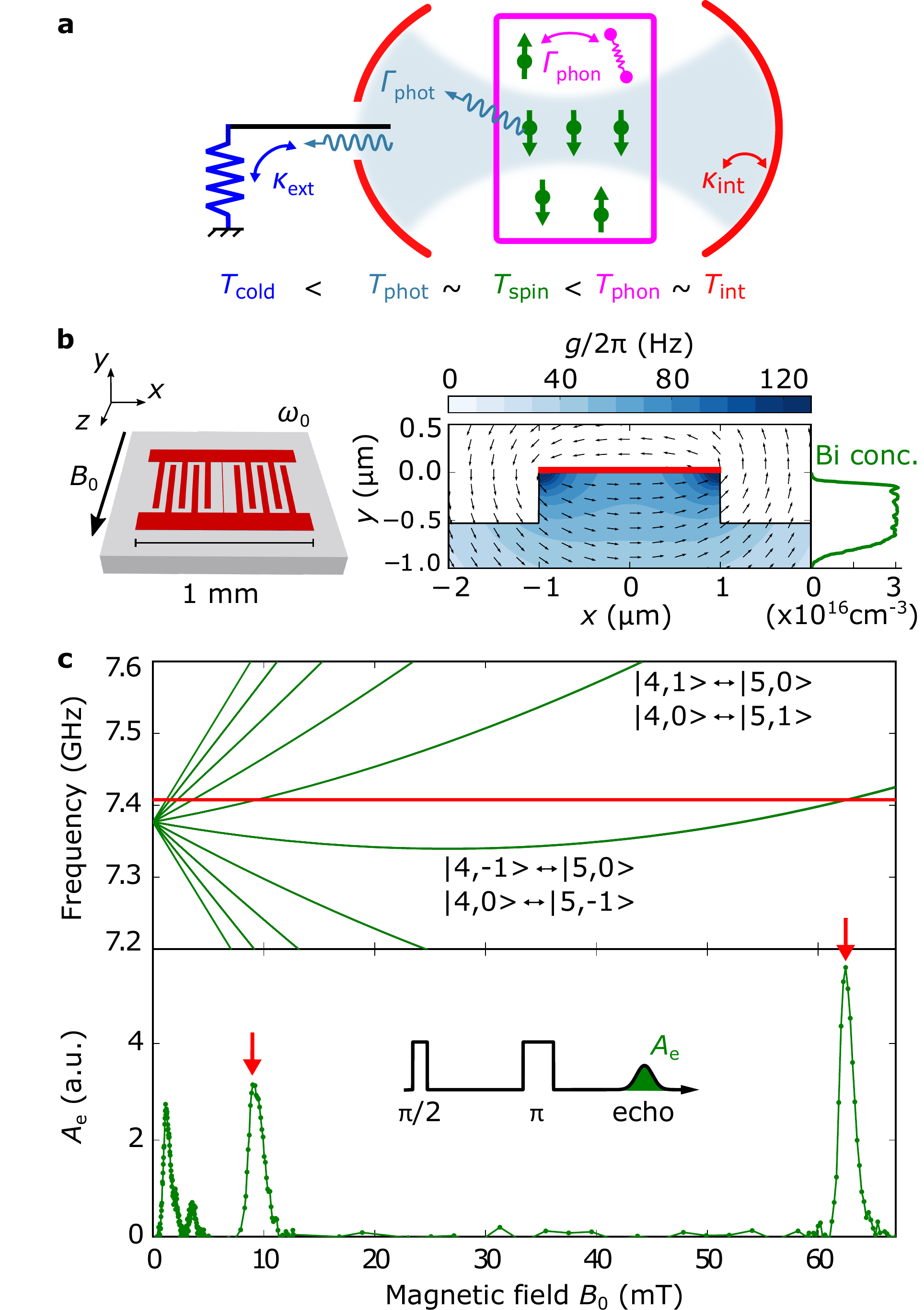}
  \caption{\label{fig1}
  \textbf{Radiative spin cooling principle, and energy spectrum of Bi donors in silicon. a,} Principle of the experiment. Spins (green) in a crystal (magenta) are coupled both to a bath of phonons at temperature $T_\text{phon}$ with a rate $\Gamma_\text{phon}$, and to a bath of microwave photons at a temperature $T_\text{phot}$ with a rate $\Gamma_\text{phot}$, which determines their equilibrium temperature $T_\text{spin}$. The temperature of the photons $T_\text{phot}$ is in turn determined by their coupling with rate $\kappa_\text{int}$ to the cavity internal losses at temperature $T_\text{int}$ and with rate $\kappa_\text{ext}$ to the load located at the cavity input. When this load is placed at a low temperature $T_\text{cold}$, the intra-cavity field is radiatively cooled provided $\kappa_\text{ext} \gg \kappa_\text{int} $, and the spins are cooled in turn if $\Gamma_\text{phot} \gg \Gamma_\text{phon}$. \textbf{b,} Left, a 50\,nm thick niobium superconducting resonator (red) of frequency $\omega_0$ is fabricated on a silicon chip where bismuth atoms are implanted. A static magnetic field $B_0$ is applied parallel to the central inductor wire of the resonator. Right, device cross section around the inductor wire (red). The spin-photon coupling constant $g$ (color code) is represented in the Bi-doped region together with the normalized $B_1$ field vector (arrows) generated by the inductor current. \textbf{c,} Top, Solid green lines are calculated electron spin resonance transitions of the bismuth donors. The resonator frequency $\omega_0/2\pi =7.408$\,GHz (red line) is resonant with 6 transitions in the 0-70\,mT range. (Bottom) Measured spin-echo amplitude $A_\text{e}$ as a function of $B_0$ (blue dots), showing the expected transitions.
  }
\end{figure}

The strength of radiative relaxation can however be enhanced by coupling the spins resonantly to one mode of a microwave resonator of frequency $\omega_0=\omega_\text{spin}$, as discovered by Purcell \cite{purcell_spontaneous_1946,butler_polarization_2011,wood_cavity_2014}. The coupling strength is then given by $\Gamma_\text{phot} = 4 g^2 / \kappa$, $g$ being the spin-photon coupling constant and $\kappa$ the resonator mode relaxation rate. Superconducting micro-resonators can be designed with a small mode volume, which increases $g$, while retaining a high quality factor, which reduces $\kappa$. This makes it possible to reach the {\it Purcell regime} defined by $\Gamma_\text{phot} \gg \Gamma_\text{phon}$, as demonstrated in recent experiments \cite{bienfait_controlling_2016}. In this regime, $T_\text{spin}$ should thus be equal to $T_\text{phot}$ and no longer to $T_\text{phon}$, and electron spin hyperpolarisation should be possible by cooling the microwave field down to a temperature $T_\text{phot} \ll T_\text{phon}$. The simplest way to do so is to connect the resonator input to a cold $50\,\Omega$ resistor, as shown in Fig.~\ref{fig1}a. As long as the coupling rate of the resonator field $\kappa_\text{ext}$ to this cold reservoir at temperature $T_\text{cold}$ is larger than the coupling rate $\kappa_\text{int} = \kappa - \kappa_\text{ext}$ to the hot reservoir consisting of its internal losses at temperature $T_\text{int}$, the resonator field temperature is lowered, which in turn cools down the spins (see Fig.\ref{fig1}a).
To infer $T_\text{spin}$, we use the fact that the amplitude of a spin-echo is proportional to $(N_\uparrow-N_\downarrow) \equiv \Delta N$, $N_\uparrow$ and $N_\downarrow$ being the number of spins in the excited and ground states, respectively. For an ensemble of $N = N_\uparrow + N_\downarrow$ spin-1/2 systems, $\Delta N = N p(T_\text{spin})$ with $p(T_\text{spin}) = \tanh{(\hbar\omega_\text{spin}/2 k T_\text{spin})}$ the temperature-dependent spin polarisation~\cite{abragam_principles_1961}.

Another key measurable quantity is the rate $\Gamma_1 $ at which the spins return to thermal equilibrium. In the Purcell regime, this rate $\Gamma_1 =\big[2\bar{n}(T_\text{phot}) + 1 \big] \Gamma_\text{phot}$ is expected to be temperature-dependent because the rate of absorption and stimulated emission of microwave photons by each spin depends on the average intra-resonator thermal photon number $\bar{n}(T_\text{phot}) = 1/[1 - e^{-\hbar \omega_0 / k T_\text{phot}}]$ ~\cite{einstein_a._strahlungs-emission_1916,haroche_exploring_2006,butler_polarization_2011}. Noting that $1/\big[2\bar{n}(T_\text{phot}) + 1 \big] = \tanh{(\hbar\omega_\text{spin}/2 k T_\text{phot})}$ we thus expect $\Gamma_1(T_\text{phot})/\Gamma_1(0) = p(0)/p(T_\text{phot})=\big[2\bar{n}(T_\text{phot}) + 1 \big]$ in the Purcell regime, an interesting prediction that we test in this work.

This radiative spin cooling scheme requires that across the temperature range between $T_\text{cold}$ and $T_\text{phon}$, spins are in the Purcell regime ($\Gamma_\text{phot} \gg \Gamma_\text{phon}$), and the resonator is over-coupled ($\kappa_\text{ext} \gg \kappa_\text{int}$). Here we use the electron spin of donors in silicon, which have low spin-lattice relaxation rate $\Gamma_\text{phon} \sim 10^{-4} - 10^{-3}\,s^{-1}$ below $1$\,K at $5-10$\,GHz\,\cite{feher_electron_1959-1,tyryshkin_electron_2012}, coupled to a superconducting micro-resonator in niobium, whose internal losses $\kappa_\text{int}$ are very low in the same temperature interval, to demonstrate radiative spin cooling from $T_\text{phon}\sim$ 0.85\,K using a cold load at $T_\text{cold}=20$\,mK.

The device geometry is depicted in Fig.~\ref{fig1}c. The niobium resonator consists of an inter-digitated capacitor shunted by a 2\,$\mu m$-wide inductive wire. It is patterned on top of a silicon substrate in which bismuth atoms were implanted over a $\sim 1\,\mu \text{m}$ depth. A dc magnetic field $B_0$ is applied parallel to the sample surface, along the inductor ($z$ axis), to tune the spin resonance by Zeeman effect. The resonator inductance generates a spatially inhomogeneous microwave magnetic field $B_1$, which results in a spatial dependence of the spin-resonator coupling constant $g$ shown in Fig.~\ref{fig1}b (see also Methods). This leads to a variation of the spin rotation angles and also relaxation rates $\Gamma_\text{phot}$ throughout the sample, which are taken into account in the simulations discussed below. 

The donor spin Hamiltonian $H =  - \gamma_e B_0 S_z + A \bold{S} \cdot \bold{I}$ describes the Zeeman shift of the $S=1/2$ electron spin of the donor with a gyromagnetic ratio $\gamma_e/2\pi = 28$\,GHz/T and its hyperfine interaction of strength $A/2\pi = 1.45$\,GHz with the nuclear spin $I=9/2$ of the bismuth atom~\cite{feher_electron_1959,mohammady_bismuth_2010}. In the low-field regime $\gamma_e B_0 \lesssim A$, the $20$ energy eigenstates are hybridized electro-nuclear spin states, characterized by the eigenvalues $F$ of the total angular momentum $\bold{F} = \bold{I}+\bold{S}$ and $m$ of its projection along $z$, $F_z$. They can be grouped into a $F=4$ ground state manifold of $9$ states and a $F=5$ excited state manifold of $11$ states, separated by $\sim 7.38$\,GHz (see Methods). All $\Delta m = \pm1$ transitions are allowed in this regime; the dependence of their frequency on $B_0$ is shown in Fig.~\ref{fig1}c. Corresponding peaks are observed in an echo-detected field sweep (see Fig.~\ref{fig1}c). Data in the following are obtained either at $B_0 = 62.5$\,mT on the quasi-degenerate $\ket{F=4,m=-1}\leftrightarrow\ket{5,0}$ and $\ket{4,0}\leftrightarrow\ket{5,-1}$ transitions, or at $B_0 = 9.5$\,mT on the seemingly quasi-degenerate $\ket{4,1}\leftrightarrow\ket{5,0}$ and $\ket{4,0}\leftrightarrow\ket{5,1}$ transitions, and we treat each one in the following as an approximate effective spin-1/2 system (see Methods). 

\begin{figure}[tbph]
  \includegraphics[width=8.2cm]{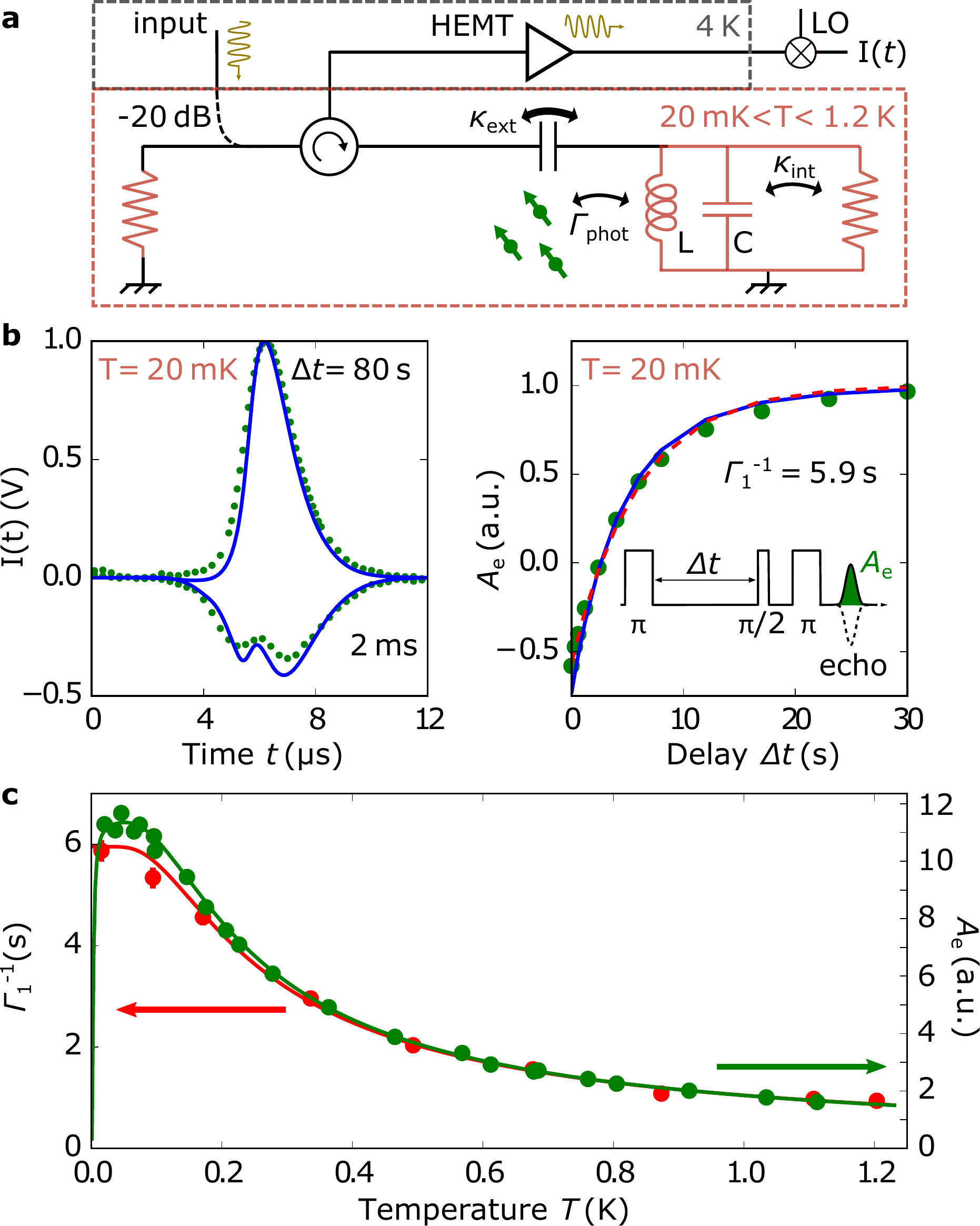}
  \caption{\label{fig2}
  \textbf{ Temperature dependence of spin relaxation rate and polarization. a,} Setup schematics. The spins, the LC resonator to which they are coupled, and the cold load are all thermalized to the same temperature $T$. Microwave pulses are sent to the cavity input via a $-20$\,dB coupler. The reflected signal, with the spin-echo, is amplified at $4$\,K by a High-Electron-Mobility-Transistor amplifier (HEMT) and demodulated at room-temperature, yielding the time-dependent quadrature $I(t)$. \textbf{b,} Measured (green dots) and simulated (blue line) echo signals (left panel) at short and long $\Delta t$, as well as integrated echo $A_e$ (right panel) as a function of $\Delta t$, both at $T=20$\,mK and $B_0=62.5$\,mT. The pulse sequence is shown in the right panel. The red dashed curve is an exponential fit with $\Gamma_1^{-1}=5.9$\,s. \textbf{c,} Measured relaxation time $\Gamma_1^{-1}$ (red dots) and integrated echo $A_\text{e}$ (green dots) as a function of temperature $T$, at $B_0=62.5$\,mT. The detailed measurement procedure is explained in the Methods. Red curve is a fit to the $\Gamma_1^{-1}(T)$ data between $0.3$\,K and $1$\,K using the function $\Gamma_{f}^{-1}/[2 \bar{n}(T)+1]$, with $\Gamma_{f}^{-1}$ as adjustable parameter. Green curve is a fit to the $A_\text{e}(T)$ data between $0.3$\,K and $1$\,K using the function $A_\text{f}\Delta N(T)$, where $\Delta N(T)$ is the expected population difference for the transition considered and $A_\text{f}$ is the adjusted parameter. The green curve deviates from the red below $0.2$\,K due to the $20$ levels of the bismuth donor system.
  }
\end{figure}

In a first experiment, we measure the spin polarization $p$ and relaxation time $\Gamma_1^{-1}$ as a function of the temperature $T$, in a setup ensuring that $T_\text{phot} = T_\text{phon} = T_\text{spin} =T_\text{int} = T$ (see Fig.~\ref{fig2}a). As shown in Fig.\,\ref{fig2}b, we first apply a $\pi$ pulse to excite the spins, let them relax during a varying time $\Delta t$, and finally measure a Hahn echo whose integral $A_e$ is proportional to $\Delta N (\Delta t)= N p(T_\text{spin}) (1 - 2 e^{- \Gamma_1 \Delta t})$, a pulse sequence known as inversion recovery~\cite{schweiger_principles_2001}. Echo curves are shown in Fig.\,~\ref{fig2}b for $\Delta t \ll \Gamma_1^{-1}$ and $\Delta t \gg \Gamma_1^{-1}$, measured at $T=20$\,mK. The phase of the short-$\Delta t$ echo is inverted with respect to the long-$\Delta t$ echo, as expected. The shape of the short-$\Delta t$ echo is due to the dependence of the Rabi angle on detuning and is well reproduced by simulations (see Fig.~\ref{fig2}b). Figure~\ref{fig2}b shows $A_e(\Delta t)$ at $20$\,mK. The curve is well fitted by an exponential decay of time constant $\Gamma_1^{-1}=5.9$\,s (dashed curve), and in excellent agreement with a complete simulation of the experiment that considers Purcell relaxation without any adjustable parameter (solid green curve), confirming that the donors are in the Purcell regime at $20$\,mK. We measure $A_e(\Delta t)$ for varying temperatures $T$; an exponential fit then yields both the relaxation time $\Gamma_1^{-1}(T)$ as well as the steady-state spin-echo area $A_e(T)$. Both quantities are shown in Fig.~\,\ref{fig2}c, with a scale evidencing their similar temperature dependence, in quantitative agreement with the expected $1/[2 \bar{n}(T)+1]$ dependence. This demonstrates that the bismuth donor spins are in the Purcell regime at least up to $1$\,K.

\begin{figure}[tbph]
  \includegraphics[width=8.2cm]{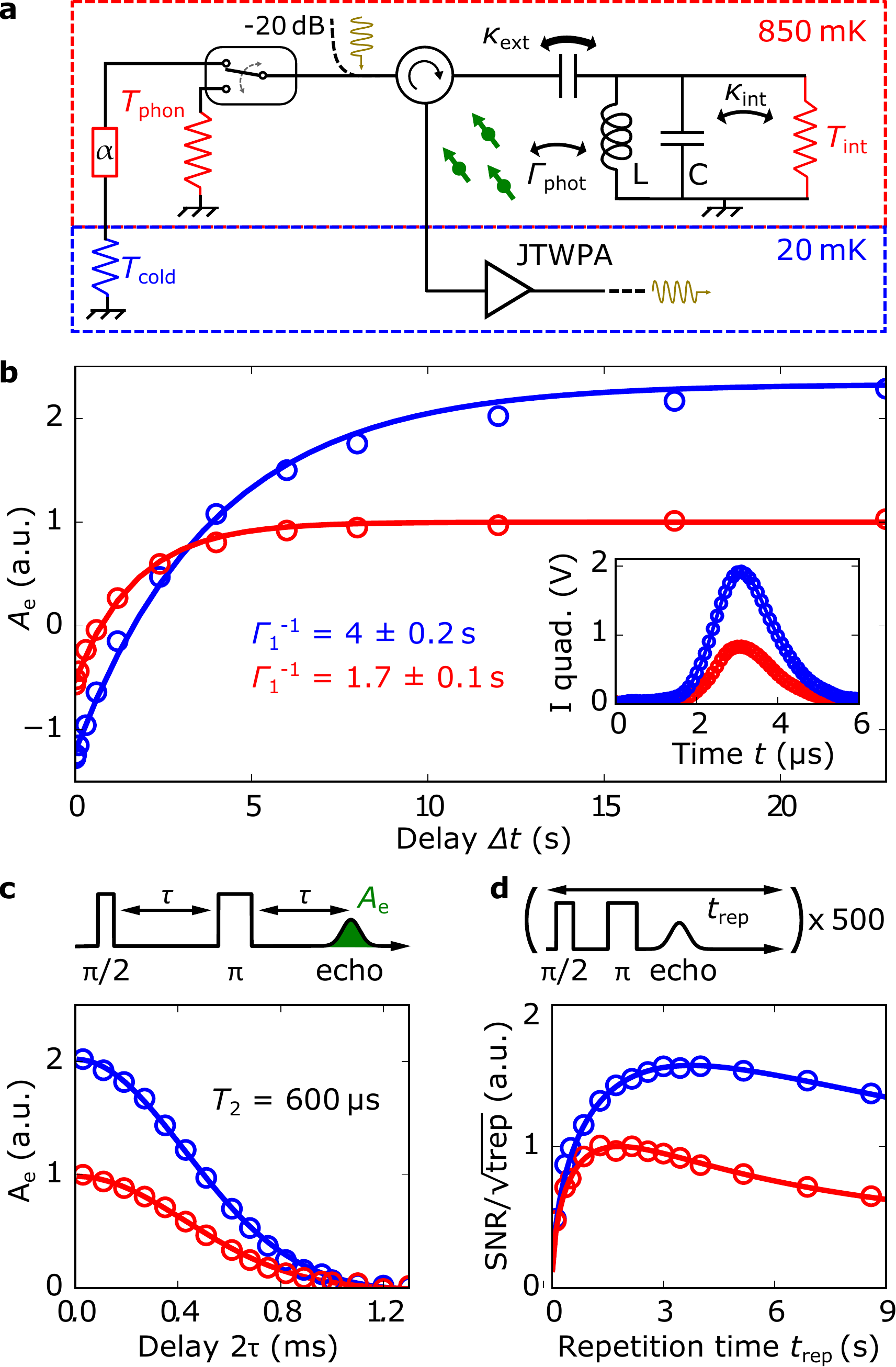}
  \caption{\label{fig3}
  \textbf{Radiative cooling demonstration a.} Simplified equivalent electrical circuit. The spins and resonator are thermally anchored at $T_\text{phon}=0.85$\,K. The resonator input is connected to an electromechanical switch that can either be connected to a hot ($T_\text{phon}$) or cold ($T_\text{cold}$) $50\,\Omega$ load. The output signal is routed via a circulator towards a Josephson Traveling-Wave Amplifier (JTWPA) installed at $20$\,mK. Excitation microwave pulses are sent via a -20\,dB directional coupler. \textbf{b,} (inset) Measured spin-echo signal showing a $\eta=2.3$ increase in amplitude in the {\it cold} configuration. (Main panel) Measurements (open circles) and exponential fits (solid lines) of the integrated echo area $A_\text{e}$ as a function of the waiting time $\Delta t$ of the inversion recovery pulse sequence shown in Fig.~\ref{fig2}b, yielding $\Gamma_1^{-1}=4\pm 0.2$\,s ({\it cold}) and $\Gamma_1^{-1}=1.7\pm 0.1$\,s ({\it hot}). \textbf{c,} Measurements (open circles) and Gaussian fits (solid lines) of the integrated echo $A_\text{e}$ as a function of waiting time $\tau$ in a Hahn echo sequence (top), yielding $T_2 = 600\,\mu \mathrm{s}$ in both switch configurations. \textbf{d,} Measured (open circles) signal-to-noise ratio SNR, obtained by dividing the mean value of 500 echo samples by their standard deviation, as a function of the repetition time $t_\text{rep}$ (see pulse sequence on top), for both switch configurations. Solid lines are fit with $p(1-e^{-\Gamma_1}t_\text{rep})/(\bar{\sigma}\sqrt{t_\text{rep}})$, where $\bar{\sigma}$ is the mean of all the $\sigma_e$ and $p$ is the equilibrium polarization, yielding $\eta = \Gamma_1^\text{hot}/\Gamma_1^\text{cold} = p^\text{cold}/p^\text{hot} = 2.1$. In all panels red and blue represent data or models in the $hot$ and $cold$ switch configuration, respectively.
  }
\end{figure}

The setup for radiative spin cooling is depicted in Fig.~\ref{fig3}a. The device is now thermally anchored at the still stage of the dilution refrigerator, so that $T_\text{phon} = 0.85$\,K. We use an electromechanical switch to connect the resonator input, via a superconducting coaxial cable, to a 50\,$\Omega$ resistor thermalized either also at the still stage (\textit{hot} configuration), i.e. at $T_\text{phon}$ or at the mixing chamber of the cryostat (\textit{cold} configuration, $T_\text{cold} = 20$\,mK). The field leaking from or reflected onto the cavity is routed via a circulator (also anchored at the still temperature) towards the detection chain where it is first amplified by a Josephson Travelling-Wave Parametric Amplifier\,~\cite{macklin_nearquantum-limited_2015}. Microwave control pulses are sent into the cavity via a $20$\,dB directional coupler. This ensures that the temperature of the microwave radiation field incident onto the resonator, and therefore $T_\text{phot}$, is dominantly determined by the switch setting. 

We first estimate the temperature $T_\text{phot}^\text{cold}$ of the intra-cavity field when the switch is connected to the cold load, using a series of noise measurements described in the Methods section, similar to recent work~\cite{wang_quantum_2019,xu_radiative_2019}. We obtain $T_\text{phot}^\text{cold} = 500 \pm 60$\,mK, proving that the microwave field is indeed cooled radiatively, but that this cooling is only partial, which we attribute to the presence of microwave losses $\alpha$ in-between the cold load and the resonator at the temperature $T_\text{phon}$ (see Methods). 

Hahn-echoes are then measured for the two different switch settings, under otherwise identical conditions (drive pulse amplitude, and repetition time). As seen in Fig.~\ref{fig3}b, the echo amplitude more than doubles when the resonator is connected to the cold load, which demonstrates radiative spin hyperpolarization with $\eta \equiv p^\text{cold}/p^\text{hot} = 2.3 \pm 0.1$. We also measure $\Gamma_1^{-1}$ in the two switch configurations, and find that $\Gamma_1^\text{hot}/\Gamma_1^\text{cold} = \eta$, as expected. Contrary to the energy relaxation time, the coherence time $T_2$ is not affected by the switch settings, as demonstrated in Fig.~\ref{fig3}c which shows $A_e(2\tau)$ for both configurations. The decay is Gaussian in both cases, with a characteristic time $T_2 = 600\,\mu \text{s}$, due to spectral diffusion in the bath of $^{29}\text{Si}$ nuclear spins~\cite{feher_electron_1959,george_electron_2010}. This confirms that the change in echo amplitude in Fig.~\ref{fig3}b is due to enhanced spin polarization, rather than to a change in $T_2$.

The measured cooling efficiency $\eta = 2.3$ corresponds to a spin temperature $T_\text{spin}^\text{cold}=350\pm10$\,mK, which is close to (or even slightly lower than) the field temperature $T_\text{phot}^\text{cold}$ estimated from noise measurements. This proves that spin-lattice relaxation is not the factor limiting the spin cooling, and that the spins thermalize to the temperature of their electromagnetic environment within the accuracy of the experiment. The spin temperature achieved is explained by $\alpha = 0.23\pm0.03$, corresponding to $1.15\pm0.15$\,dB loss at $850$\,mK, a plausible value for the combined effect of circulator insertion losses, directional coupler contribution, and possible spurious reflections due to impedance mismatch in the line. 

The sensitivity enhancement obtained by radiative spin hyperpolarization does not scale like $\eta$ as in other hyperpolarization schemes such as DNP but only as $\sqrt{\eta}$, because the optimal waiting time $t_\text{rep}$ between subsequent experimental sequences, which is of order $\Gamma_1^{-1}$, also scales as $\eta$, as seen above. We demonstrate this by measuring the mean value and standard deviation of $A_e$ for $500$ echo traces as a function of the repetition time $t_\text{rep}$ (see Fig.~\ref{fig3}d). The highest sensitivity is obtained for $t_\text{rep} \simeq 1.25 [\Gamma_1^\text{cold,hot}]^{-1}$, both in the cold and the hot load cases. It is $1.6$ times larger in the $cold$ than in the $hot$ configuration, slightly larger than $\sqrt{\eta} = 1.52$ because switching to the cold load also substantially reduces the effective noise temperature of our detection chain (by $7\%$).

The presence of an additional non-radiative coupling of the spins with rate $\Gamma_\text{phon}$ to a bosonic bath of temperature $T_\text{phon}$ should enter in competition with radiative cooling, and therefore reduce its efficiency to $\eta = \Gamma_1^\text{hot} / \Gamma_1^\text{cold}$, with $\Gamma_1^\text{cold,hot}=\Gamma_\text{phon}[2\bar{n}(T_\text{phon})+1]+\Gamma_\text{phot}[2\bar{n}(T_\text{phot}^\text{cold,hot})+1]$.

\begin{figure}[tbph]
  \includegraphics[width=8.2cm]{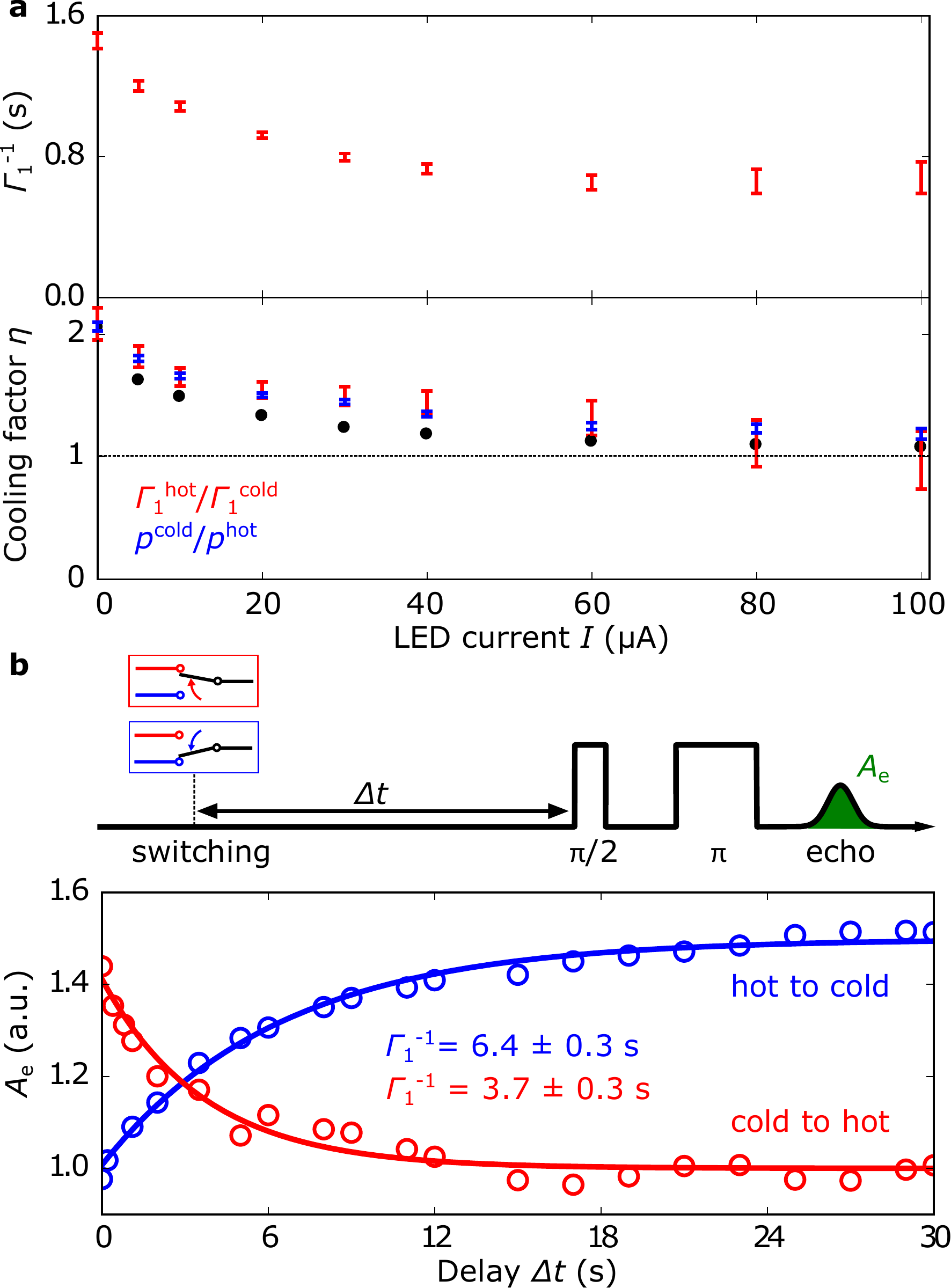}
  \caption{\label{fig4}
  \textbf{Suppression of cooling by carrier injection, and cooling dynamics. a,} (Top) Spin relaxation time $\Gamma_1^{-1}$, measured with an inversion recovery sequence in the $hot$ configuration, as a function of the current $I$ through the LED. (Bottom) Measured cooling factor $\eta$ (blue) and $\Gamma_1^\text{hot}/\Gamma_1^\text{cold}$ (red) as a function of $I$. The black dots are the prediction of the model described in the main text. Error bars are estimated from the standard deviation of the exponential fit parameters. \textbf{b,} Measured integrated echo area $A_\text{e}$ (open circles) at 9.5\,mT as a function of the waiting time $\Delta t$ between a rapid change of switch configuration and the echo sequence shown on top. Solid lines are exponential fits, yielding $\Gamma_1^{-1}=6.4\pm 0.3$\,s from hot to cold (blue) and $\Gamma_1^{-1}=3.7 \pm 0.3$\,s from cold to hot (red).
  }
\end{figure}

We test this prediction by controllably exciting carriers into the silicon conduction band, which are known to reduce the spin relaxation time of donors~\cite{feher_electron_1959-1}. For this we irradiate the sample with infrared radiation at $950$\,nm emitted by a LED biased by a current $I$. The spin relaxation time $\Gamma_1^{-1}$, measured in the $hot$ configuration, is indeed observed to decrease with $I$ (see Fig.~\ref{fig4}a), from which we determine $\Gamma_\text{phon}^{-1}(I)$. We then measure the cooling efficiency $\eta(I)$ (see Fig.~\ref{fig4}a). At large LED powers, radiative spin cooling stops working as expected ($\eta \sim 1$), because spins relax dominantly by interaction with the charge carriers in the substrate, whose presence is unaffected by the switch settings. The $\eta(I)$ dependence predicted by our model agrees semi-quantitatively with the data. The ratio of relaxation times $\Gamma_1^\text{hot}(I)/\Gamma_1^\text{cold}(I)$ also closely follows the measured $\eta(I)$, as expected.

We finally investigate the spin polarization dynamics (see Fig.~\ref{fig4}b). For that we replace the electromechanical switch with a superconducting device mounted at 20\,mK, able to switch in a few nanoseconds without heating~\cite{pechal_superconducting_2016}. At 9.5\,mT the area $A_e$ of a Hahn echo is measured at delay $\Delta t$ after the switch configuration is changed, either from {\it cold} to {\it hot} or vice-versa. The relaxation times in the two cases are expected to be the two values of $\Gamma_\text{phot}[2\bar{n}(T^\text{cold,hot})+1]$ determined by the new thermal photon population after the switching. An exponential fit of the two curves gives $(\Gamma_{1}^\text{hot})^{-1} =3.7$\,s and $(\Gamma_{1}^\text{cold})^{-1} =6.4$\,s. The approximate equality of the two measured ratios, $\Gamma_1^\text{hot}/\Gamma_1^\text{cold} =1.7$ and $p^\text{cold}/p^\text{hot} =1.5$, is again in agreement with our model. The reduced $\eta$ with respect to the previously presented result is explained by a measured 3dB insertion loss of the superconducting switch.

We now come to the limitations and potential real-world applications of our scheme. Cooling the spins while keeping the sample hot is useful in situations where large cooling powers are needed, for instance in spin-based microwave-optical transduction experiments which require large optical powers~\cite{williamson_magneto-optic_2014}. One limitation of the present scheme is that the lowest spin temperature that could possibly be reached (in a lossless setup) is the temperature $T_\text{cold}$ at which the cold load is thermally anchored. However, one could also imagine cooling the resonator field using parametric processes in a circuit quantum electrodynamics platform, as recently demonstrated in~\cite{gely_observation_2019}. This would enable to cool the resonator field and thus the spins to arbitrarily low temperatures, an appealing perspective for magnetic resonance. Such an active radiative cooling scheme could work even in $^4$He cryostats with base temperature 1.5-4\,K, as in standard low-temperature EPR spectrometers. Even higher operating temperatures could be envisioned by using high-Tc superconducting thin-films, although the microwave losses caused by quasiparticles may be an issue~\cite{rauch_microwave_1993}.

Regarding applications to magnetic resonance spectroscopy, we first note that to our knowledge the only other existing electron spin hyperpolarisation method is optical illumination, which requires spin systems with very specific level schemes~\cite{adrian_theory_1971,wong_chemically_1973,steger_quantum_2012,doherty_nitrogen-vacancy_2013}, contrary to our method which only makes use of the spin transition. A broad set of Electron Paramagnetic Resonance (EPR) measurements (including field-sweeps, g-tensor measurements, HYSCORE, DEER and ENDOR etc.) could benefit from an increase in sensitivity of at least $\sqrt{\eta}$, translating into a measurement time shorter by a factor of $\eta$, obtained by radiative hyperpolarisation. Only those EPR studies where the spin-lattice relaxation rate is itself the object of interest would be excluded from this approach. Regarding suitable spin species, the key requirement is for the spins to be in the Purcell regime. At cryogenic temperatures, electron spin-lattice relaxation times $\Gamma_\text{phon}^{-1}$ are typically in the $10^{-3}-10^3$\,s range~\cite{castle_resonance_1965,gayda_temperature_1979,zhou_electron_1999}, so that a Purcell time $\Gamma_\text{phot}^{-1} \sim 10^{-2}$\,s, which has already been demonstrated with a different resonator geometry~\cite{probst_inductive-detection_2017}, would be sufficient for many species. The method could also be fruitfully applied to the radicals used in DNP as polarizing agents, and might enable to obtain large nuclear spin enhancement factors in less demanding conditions than those usually required~\cite{ardenkjaer-larsen_increase_2003} (lower magnetic field, lower microwave frequency, or higher temperature).

We have shown that in the Purcell regime, the spin temperature is determined by the temperature of the microwave field and not of the sample itself. We have used this fundamental insight to demonstrate a novel electron spin hyperpolarisation method, which is effective for all spin systems provided they can reach the Purcell regime. Further work will aim at demonstrating active Purcell cooling, and investigate real-world applications of our method. 

\textbf{Acknowledgements} We thank P.~S\'enat, D. Duet and J.-C. Tack for the technical support, and are grateful for fruitful discussions within the Quantronics group. We acknowledge IARPA and Lincoln Labs for providing a Josephson Traveling-Wave Parametric Amplifier. We acknowledge support of the European Research Council under the European Community's Seventh Framework Programme (FP7/2007-2013) through grant agreement No.~615767 (CIRQUSS), of the Agence Nationale de la Rercherche under the Chaire Industrielle NASNIQ, of the Région Ile-de-France via the DIM SIRTEQ, of the Engineering and Physical Sciences Research Council (EPSRC) through Grant No. EP/ K025945/1, and of the Horizon 2020 research and innovation programme through grant agreement No. 771493 (LOQO-MOTIONS).  

\textbf{Author contributions} B.A., S.P. and P.B. designed the experiment. J.J.L.M. and C.Z. provided and characterized the implanted silicon sample, on which B.A. and S.P. fabricated the niobium resonator. B.A. performed the measurements, with help from S.P. and V.R.  B.A. and P.B. analyzed the data. B.A. and V.R. performed the simulations. M.P. realized and tested the superconducting switch in a project guided by A.W.  B.A. and P.B. wrote the manuscript. S.P., V.R., A.W., J.J.L.M, D.V., D.E., and E.F. contributed with useful input to the manuscript.

\bibliographystyle{unsrt}
\bibliography{Albanese}
\newpage
\textbf{}
\newpage
\textbf{SUPPLEMENTARY INFORMATION}
\newline

\textbf{Thermalization of a system coupled to $N$ bosonic baths.} 
When a physical system exchanges energy with an environment consisting of $N$ reservoirs at different temperatures $T_\text{j}$, the intermediate effective temperature $T_\text{sys}$ at which the system equilibrates depends on the strength $\Gamma_j$ with which it is coupled to each bath, defined as the rate at which the system would spontaneously relax from its first excited to its ground state by emitting a quantum of energy into this bath at zero temperature. For a system coupled to $N$ bosonic reservoirs, $T_\text{sys}$ is obtained by 

\begin{equation}
\label{eqMethods:Tsys}
\bar{n}(T_\text{sys})  =  \sum_{j=1}^{N} (\Gamma_j/\Gamma) \bar{n}(T_j),
\end{equation}

\noindent $\bar{n}(T) = 1/[1 - \exp ^{-\hbar \omega_0 / k T}]$ being the occupation number of a bosonic mode at frequency $\omega_0$, and $\Gamma = \sum_{j=1}^{N} \Gamma_j$ the total system-bath coupling. If the system is dominantly coupled to one bath $j0$ ($\Gamma_{j0} \gg \Gamma_{j \neq j0}$), the system therefore equilibrates close to $T_{j0}$ regardless of the temperature of the other reservoirs. 

\textbf{Cavity thermal photon population.} 
The resonator mode of frequency $\omega_0$ is coupled by internal dissipation of rate $\ki$ to a hot thermal bath at $\Ti$, not necessarily equal to the sample temperature $\Tphon$, and by the external losses of rate $\ke$, to the thermal radiation in the propagating input mode, whose temperature $T_\text{ext}$ is determined by the switch state. From Eq.\ref{eqMethods:Tsys} we get $\bar{n}(T_\text{phot}) = (\ki/\kappa)\bar{n}(\Ti)+(\ke/\kappa)\bar{n}(T_\text{ext})$, where $\kappa = \ki+\ke$. In the same way, when the switch is in the $cold$ configuration, the input mode is coupled to the cold bath at $\Tc$ and, as a consequence of losses and impedance mismatches modeled by the absorption factor $\alpha$, to the hot bath at $\Tphon$. The thermal occupation of the input mode at $\omega_0$ in the $cold$ state is then an average of the hot and cold bath populations expressed as $\bar{n}(T_\text{ext}^\text{cold})=(1-\alpha)\bar{n}(T_\text{cold})+\alpha\bar{n}(\Tphon)$. In the present experiment $\Tc=20$\,mK, therefore $\bar{n}(T_\text{cold}) \approx 0$. Substituting $\bar{n}(T_\text{ext}^\text{cold})$ in the expression of $\bar{n}(\Tphot)$, the average number of resonator thermal photons in the $cold$ case is found to be $\bar{n}(T_\text{phot}^\text{cold})\approx \frac{\ki}{\kappa}\bar{n}(\Ti)+ \frac{\alpha\ke}{\kappa}\bar{n}(\Tphon)$. When in $hot$, $\bar{n}(T_\text{ext}^\text{hot})=\bar{n}(\Tphon)$ and $\bar{n}(\Tphoth) = (\kappa_\text{int}/\kappa)\bar{n}(T_\text{int})+(\kappa_\text{ext}/\kappa)\bar{n}(T_\text{phon})$.

\textbf{Spin effective temperature $T_\text{spin}$ and cooling factor $\eta$.} The spins are coupled with strength $\Gamma_\text{phot}$ to the resonator mode populated by $\bar{n}(T_\text{phot})$ thermal photons, and with strength $\Gamma_\text{phon}$ to the lattice phonon mode of occupation number $\bar{n}(T_\text{phon})$. From Eq.\ref{eqMethods:Tsys}, the spin temperature is defined as $\bar{n}(T_\text{spin})=[\Gamma_\text{phon}/(\Gamma_\text{phon}+\Gamma_\text{phot})]\bar{n}(T_\text{phon})+[\Gamma_\text{phot}/(\Gamma_\text{phon}+\Gamma_\text{phot})]\bar{n}(T_\text{phot})$. As discussed in the main text for the Purcell rate, a spin coupled to a bosonic mode of temperature $T_{j}$ with strength $\Gamma_j$ relaxes at a rate $\Gamma_1 = \Gamma_{j}[2\bar{n}(T_{j})+1]$ as a consequence of absorption and emission rate dependence on the thermal population of the mode. The relaxation rate of the spins coupled to the effective bath of temperature $T_\text{spin}$ is then $\Gamma_1 = (\Gamma_\text{phon}+\Gamma_\text{phot})[2\bar{n}(T_\text{spin})+1] =\Gamma_\text{phon}[2\bar{n}(T_\text{phon})+1]+\Gamma_\text{phot}[2\bar{n}(T_\text{phot})+1]$, that is the sum of the spin-photon and spin-lattice relaxation rates, as expected. On the other hand, the polarization of the spins at temperature $T_\text{spin}$ is $p(T_\text{spin})=1/[2n(T_\text{spin})+1] = (\Gamma_\text{phon}/\Gamma_1)p(T_\text{phon})+(\Gamma_\text{phot}/\Gamma_1)p(T_\text{phot})$, so equal to the equilibrium polarizations at the two baths temperatures weighted by the corresponding relaxation rates. The cooling factor, defined as the ratio of the equilibrium polarization in the two switch states, is therefore still equal to the ratio of total relaxation rates as for purely Purcell relaxation, $\eta = \frac{\Gamma_\text{phon}(2\bar{n}(T_\text{phon})+1)+\Gamma_\text{phot}(2\bar{n}(T_\text{phot}^\text{hot})+1)}{\Gamma_\text{phon}(2\bar{n}(T_\text{phon})+1)+\Gamma_\text{phot}(2\bar{n}(T_\text{phot}^\text{cold})+1)}$. In this work we assume $\Gamma_\text{phon}$ to be negligible in the absence of infrared radiation from the LED, as confirmed by the resonator mode temperature measurement and by simulation of the Purcell decay. In the model used to fit $\eta(I)$ in Fig.~\ref{fig4}b, we neglect a possible effect of illumination on the lattice bath temperature.

\textbf{Measurement of cavity mode temperature.} Measuring the noise power spectral density $S$ in the output mode yields an estimate of the resonator mode temperature. $S(\omega,\Tphon)$ has then been measured in a 6\,MHz bandwidth around $\omega_0$ for $\Tphon$ ranging from 840\,mK to 1.16\,K in both switch configurations. A lower resonator quality factor in the $cold$ than in the $hot$ configuration, measured at $B_0=0$, points to a reduced temperature of the internal losses $\Tic<\Tih$ possibly due to cooling of the electric dipoles coupled to the resonator. We therofore consider the general case of resonator internal losses temperature $\Ti\neq \Tphon$. In the $h$ configuration then, $S^\text{hot}/G(\omega)\hbar\omega = [1-\beta(\omega)]\bar{n}(\Tphon)+\beta(\omega)\bar{n}(\Tic)+1/2+n_\text{TWPA}$, where $G(\omega)$ is the known frequency-dependent total gain of the output line, $n_\text{TWPA}$ the noise added by the TWPA and $\beta(\omega) = 4\ki\ke/(\kappa^2+4(\omega-\omega_0)^2)$ the transmission function. On the other hand, the expression of $S$ in the $cold$ case is $S^\text{cold}/G(\omega)\hbar\omega = [\alpha(1-\beta(\omega)]\bar{n}(\Tphon)+\beta(\omega)\bar{n}(\Tic)+1/2+n_\text{TWPA})$. A first fit of $S^\text{hot}(\Tphon)$ gives $n_\text{TWPA} =0.75\pm0.25$. By then fitting $S^\text{hot}(\omega)$ for each value of $\Tphon$ we obtain the corresponding $\Tih$, while $\alpha$ and $\Tic$ result from the fit of $S^\text{cold}(\omega)$. The temperature $\Ti$ of the resonator internal losses is found to be about 10\% larger (lower) than $T_\text{phon}$ values in the $hot$ ($cold$) configuration. For all $T_\text{phon}$ values we extract $\alpha \simeq 0.47\pm0.1$. The extracted parameters allow to calculate the cavity thermal photon population using $\bar{n}(\Tphoth) = (\ke/\kappa)\bar{n}(\Tphon)+(\ki/\kappa)\bar{n}(\Tih)$ and $\bar{n}(\Tphotc)= \alpha(\ke/\kappa)\bar{n}(\Tphon)+(\ki/\kappa)\bar{n}(\Tic)$ in the $hot$ and $cold$ configuration, respectively. The effective intra-cavity field temperature of $\Tphoth=850$\,mK and $\Tphotc=500\pm60$\,mK are then found at $\Tphon =840$\,mK, corresponding to a predicted cooling factor $\eta = 1.65\pm0.2$ for the spins. The measurement of $S$ therefore confirms that the spin cooling efficiency is limited by the temperature of the intra-cavity mode rather than non-radiative spin relaxation phenomena.

\begin{figure}[tbph]
  \includegraphics[width=8.2cm]{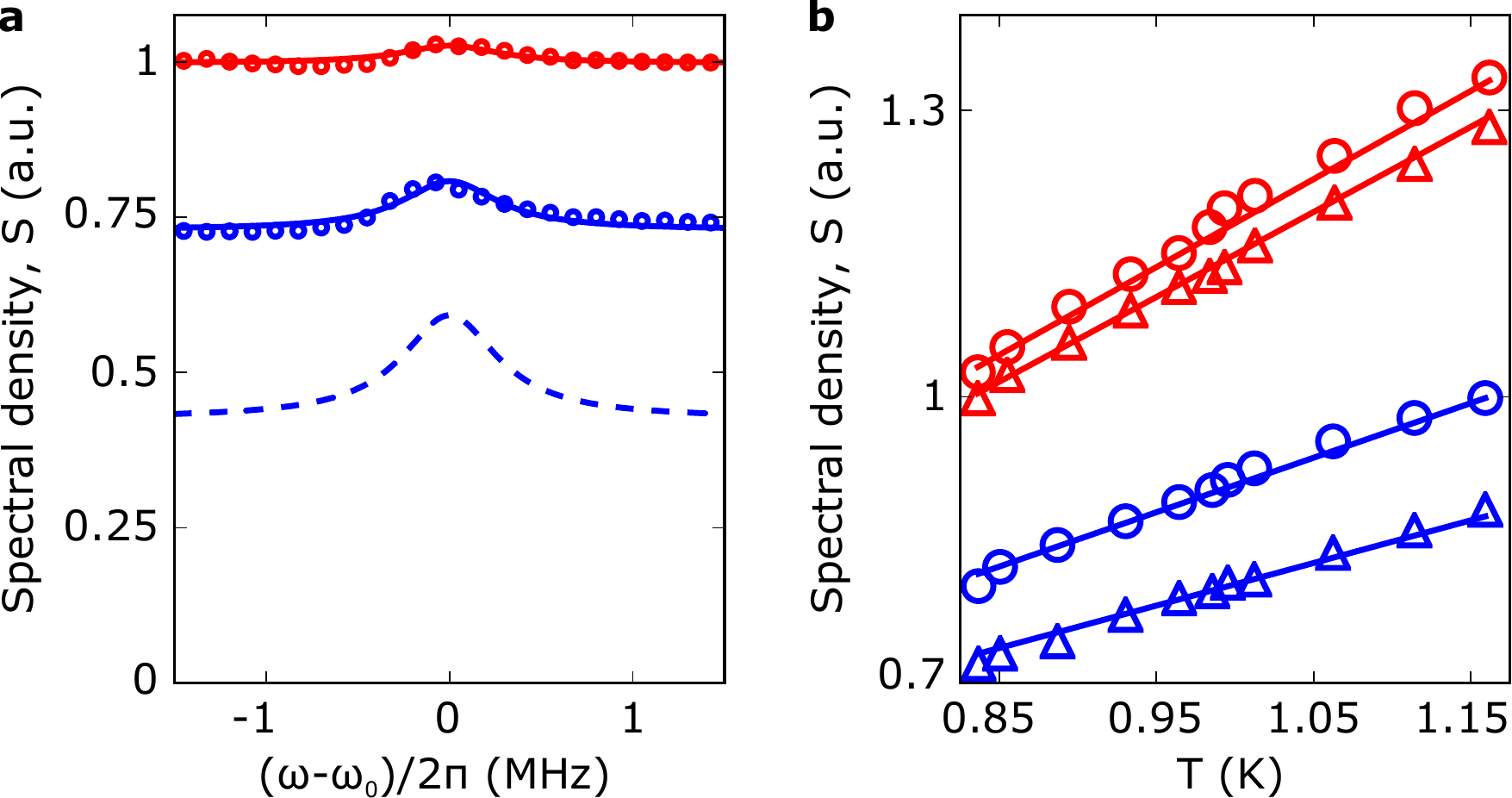}
  \caption{\label{fig1_ext}
  \textbf{Noise power spectral density measurement. a,} Frequency dependence of the noise power spectral density $S$ measured at $T_{spl} =840$\,mK for the $h$ (red circles) and $c$ (blue circles) switch configurations. Hot data are fitted with $S^{h}/\hbar\omega = G(\omega)((1-\beta(\omega))\bar{n}(T_{spl})+\beta(\omega)\bar{n}(T_i^h)+1/2+n_{TWPA})$ (solid red line), fixing $n_{TWPA}=0.75\pm0.25$ and obtaining $T_i^h =950$\,mK as the only free parameter. From the fit of the cold data with $S^{c}/\hbar\omega = G(\omega)(\alpha(1-\beta(\omega))\bar{n}(T_{spl})+\beta(\omega)\bar{n}(T_i^c)+1/2+n_{TWPA})$ (solid blue line), $\alpha=0.47\pm0.04$ and $T_{int}^{cold} =760$\,mK are obtained. The blue dashed line indicates the expected $S^{cold}(\omega)$ for $\alpha=0$. \textbf{b,} Still temperature $T_{phon}$ dependence of $S$ measured at $\omega = \omega_0$ (open circles) and at $\omega-\omega_0 =-2.7$\,MHz (open triangles) for both $hot$ (red) and $cold$ (blue) configurations. Solid lines are plot of $S^\text{hot}$ (red) and $S^{cold}$ (blue) with parameters obtained from the frequency dependence fits performed at all $T_\text{phon}$, and with $n_\text{TWPA} = 0.75$.
  }
\end{figure}

\textbf{Bismuth donors in silicon.} The spin Hamiltonian of bismuth donors in silicon includes a Zeeman effect and an isotropic hyperfine coupling: $H/\hbar = \textit{\textbf{B}}\cdot(\gamma_e \textit{\textbf{S}}\otimes\mathds{1}-\gamma_n \mathds{1}\otimes \textit{\textbf{I}})+A\textit{\textbf{S}}\cdot\textit{\textbf{I}}$, where $\gamma_e/2\pi = 27.997$\,GHz is the electronic gyromagnetic ratio, $\gamma_n/2\pi = 6.9$\,MHz is the nuclear gyromagnetic ratio and $A/h = 1.475$\,GHz is the hyperfine coupling constant. For low applied static magnetic field aligned along $z$ ($B_0\gamma_e\lesssim A$), the eigenstates of $\textit{\textbf{S}}$ and $\textit{\textbf{I}}$ are highly hybridized and the energy eigenbasis is ${\ket{F,m}}$, in which $\textit{\textbf{F}} = \textit{\textbf{S}}+\textit{\textbf{I}}$ is the total angular momentum of eigenvalue $F$ and $F_z$ its projection along $\textit{\textbf{B}}_0$ with eigenvalue $m$. An oscillating field along $x$ or $y$ can therefore excite transitions satisfying $\Delta F\Delta m = \pm 1$, whose matrix elements $\bra{F,m}S_{x}\ket{F+1,m\pm1}$=$\bra{F,m}S_{y}\ket{F+1,m\pm1}$ are comparable in magnitude with the ideal electronic spin 1/2 transition $\bra{-1/2}S_{x}\ket{1/2} = 0.5$. We note here that in the $-4<m\leq4$ manifold the $\ket{F,m-1}\leftrightarrow\ket{F+1,m}$ and $\ket{F,m}\leftrightarrow\ket{F+1,m-1}$ transitions are quasi degenerate in frequency and the sum of their associated $S_{x}$ matrix elements is equal to 0.5. The spectrum consists then of 6 resolvable transitions for $B_0<100$\,mT, as shown in Fig.~\ref{fig1}.c. At zero field the 9 ground ($F=4$) and 11 excited ($F=5$) multiplets are separated by a frequency of 7.38\,GHz. For $B_0$ going from 0 to 65\,mT, the hyperfine splitting within the two multiplets $E_\text{hyp}=E_{\ket{F,m+1}}-E_{\ket{F,m}}$ increases from 0 to about 150\,MHz.

\textbf{Population difference $\Delta N(T)$ for bismuth donors.} The magnetic resonance signal is proportional to the population unbalance $\Delta N = N_\uparrow-N_\downarrow$, where $N_\uparrow$ ($N_\downarrow$) is the total number of spins in the excited (ground) state of the spin transition considered. In the case of bismuth donors, where the electron spin transitions $\ket{4,m}\leftrightarrow\ket{5,m-1}$ and $\ket{4,m-1}\leftrightarrow\ket{5,m}$ (for $-4<m\leq4$) are quasi-degenerate in frequency compared to resonator linewidth, $\Delta N (\Ts)$ is the population unbalance between the two excited and the two ground states: $\Delta N (\Ts)=N(p_{\ket{4,m}}+p_{\ket{4,m-1}}-p_{\ket{5,m}}-p_{\ket{5,m-1}})$, where $N$ is the total number of donors, and $p_{\ket{F,m}} = {e^{{-E_{F,m}}/{k\Ts}}}/{Z}$ is the occupation probability of $\ket{F,m}$, with $Z = \sum_{F,m} e^{{-E_{F,m}}/{k\Ts}}$ the partition function. The green line in Fig.2 of the main text shows $\Delta N (T)$ for the transition considered.

As long as $\Ts>100$\,mK and $B_0<70$\,mT, the thermal energy is much larger than the energy difference between hyperfine states of both $F=5$ and $F=4$ manifolds $E_\ket{F,m} - E_\ket{F,m-1}$. In that case, one can show that $\Delta N (\Ts) \simeq \frac{N}{9}\frac{1+e^{-\hbar\omega_0/k\Ts}}{1+11/9e^{-\hbar\omega_0/k\Ts}}\tanh(\hbar\omega_0/2k\Ts)$, which can be approxmiated by $\frac{N}{10} \tanh(\hbar\omega_0/2k\Ts)$ especially when $k\Ts>\hbar \omega_0$, which happens in our case for $\Ts>300$\,mK. This is visible in Fig.~\ref{fig2}c, where the computed $\Delta N(T)$ (green curve) indeed coincides well with the result for a spin-1/2 $\tanh(\hbar\omega_0/2k\Ts)$ (red curve) for $\Ts>300$\,mK (by proper choice of the scale for $A_\text{e}$).

\begin{figure}[tbph]
  \includegraphics[width=8.5cm]{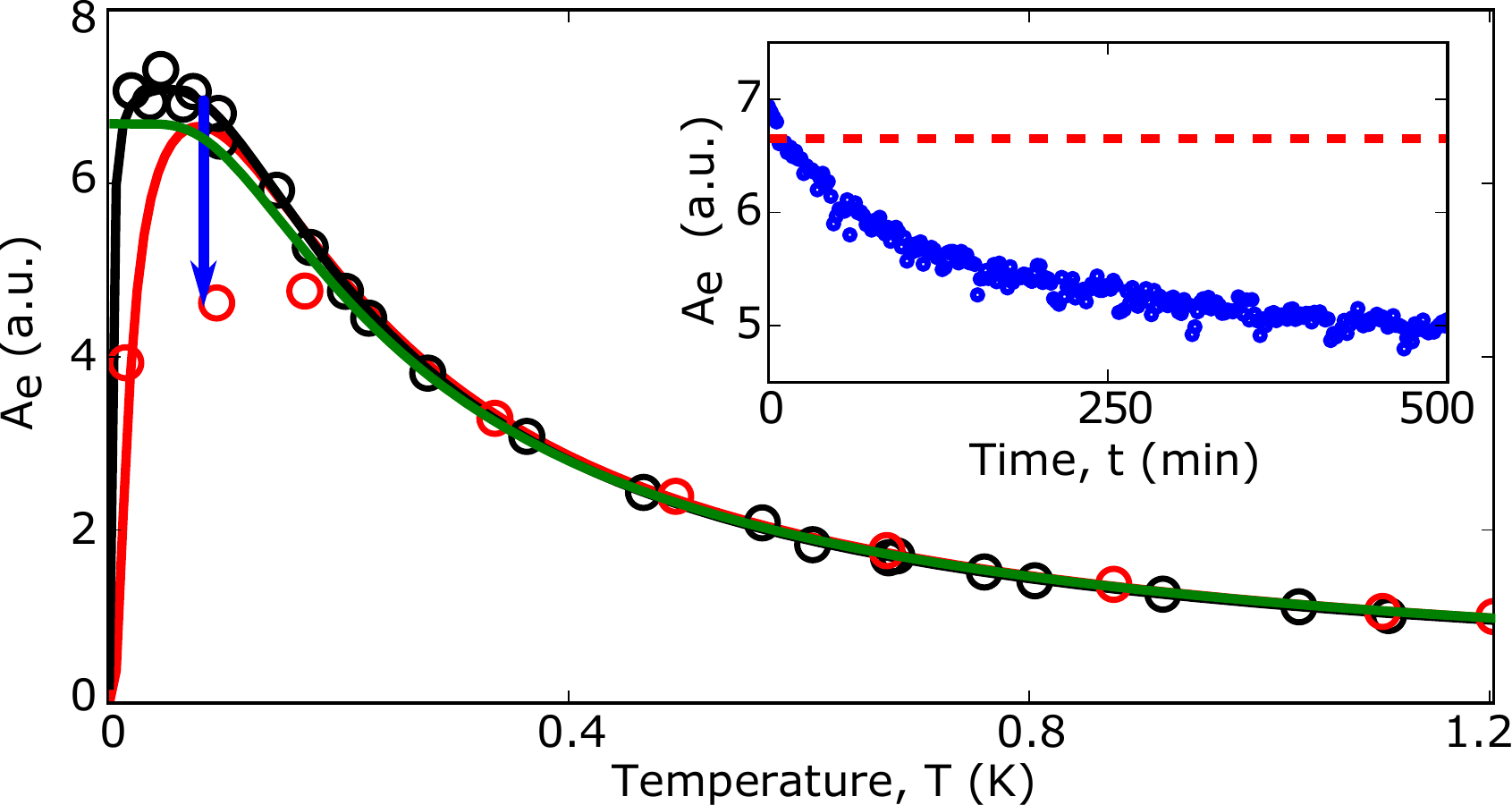}
  \caption{\label{fig2_ext}
  \textbf{Temperature dependence of polarization for transitions $\ket{4,-1}\leftrightarrow\ket{5,0}$ and $\ket{4,0}\leftrightarrow\ket{5,-1}$.} Equilibrium polarization of transitions $\ket{4,-1}\leftrightarrow\ket{5,0}$ and $\ket{4,0}\leftrightarrow\ket{5,-1}$ measured at $B_0=62.5$\,mT (red circles). Several hours are waited at each temperature before recording $A_\text{e}$. Red line is the calculated $\Delta N (T)$ for the considered transition at $B_0=62.5$\,mT. A second polarisation measurement of the same transitions (black circles) is reported. In this experiment, for each temperature value, $B_0$ is first set to 9.3\,mT during 20\,min, then it is set to 62.5\,mT and finally after 4\,min $A_\text{e}$ is recorded. The black line is the calculated $\Delta N (T)$ for the considered transition at $B_0=9.3$\,mT. The polarisation $p(T)=\tanh(\hbar\omega_0/kT)$ of a spin 1/2 is also shown for comparison (green). $A_\text{e}$ as a function of time (inset) is measured at $T=83$\,mK and $B_0=62.5$\,mT after $B_0$ has been set to 9.3\,mT for 20\,min. The same data are represented in the main plot with the blue arrow.
  }
\end{figure}

\textbf{Temperature dependence of polarization for transitions $\ket{4,-1}\leftrightarrow\ket{5,0}$ and $\ket{4,0}\leftrightarrow\ket{5,-1}$.} We first measure the temperature dependence of polarisation at $B_0=62.5$\,mT waiting several hours at each temperature value before measuring the echo amplitude $A_\text{e}$. The result reported in Fig.~\ref{fig1_ext} (red circles) shows a significant deviation below 200\,mK from the calculated $\Delta N (T)$ (red line). Spins are then in a non-thermal state below 200\,mK, possibly due to residual infrared radiation reaching the sample. This is known to cause redistribution of population in the bismuth hyperfine levels~\cite{sekiguchi_hyperfine_2010}. We repeat then the same measurement by adding at each temperature point a first step during which we set $B_0=9.3$\,mT for 20\,min before setting the field back to 62.5\,mT. 4\,min after the field is set to 62.5\,mT, we measure $A_\text{e}$. The result is reported in both Fig.~\ref{fig1}c (green dots) and Fig.~\ref{fig2_ext} (black circles) and is in good agreement with the calculated $\Delta N (T)$ for the same transitions and $B_0=9.3$\,mT. This shows that at lower field the spins reach thermal equilibrium while the depolarising phenomenon taking place at 62.5\,mT is not effective. In a third measurement performed at $T=83$\,mK we first set $B_0=9.3$\,mT during 20\,min, then we set $B_0=62.5$\,mT and immediately after we record continuously $A_\text{e}$ as a function of time. The result reported in the inset of Fig.~\ref{fig2_ext} shows the spins reaching the equilibrium non-thermal state in the time-scale of hours, orders of magnitude longer than the 4\,min waiting time used to detect the thermal equilibrium polarization after field sweep from 9.3\,mT to 62.5\,mT.

\begin{figure}[tbph]
  \includegraphics[width=8.5cm]{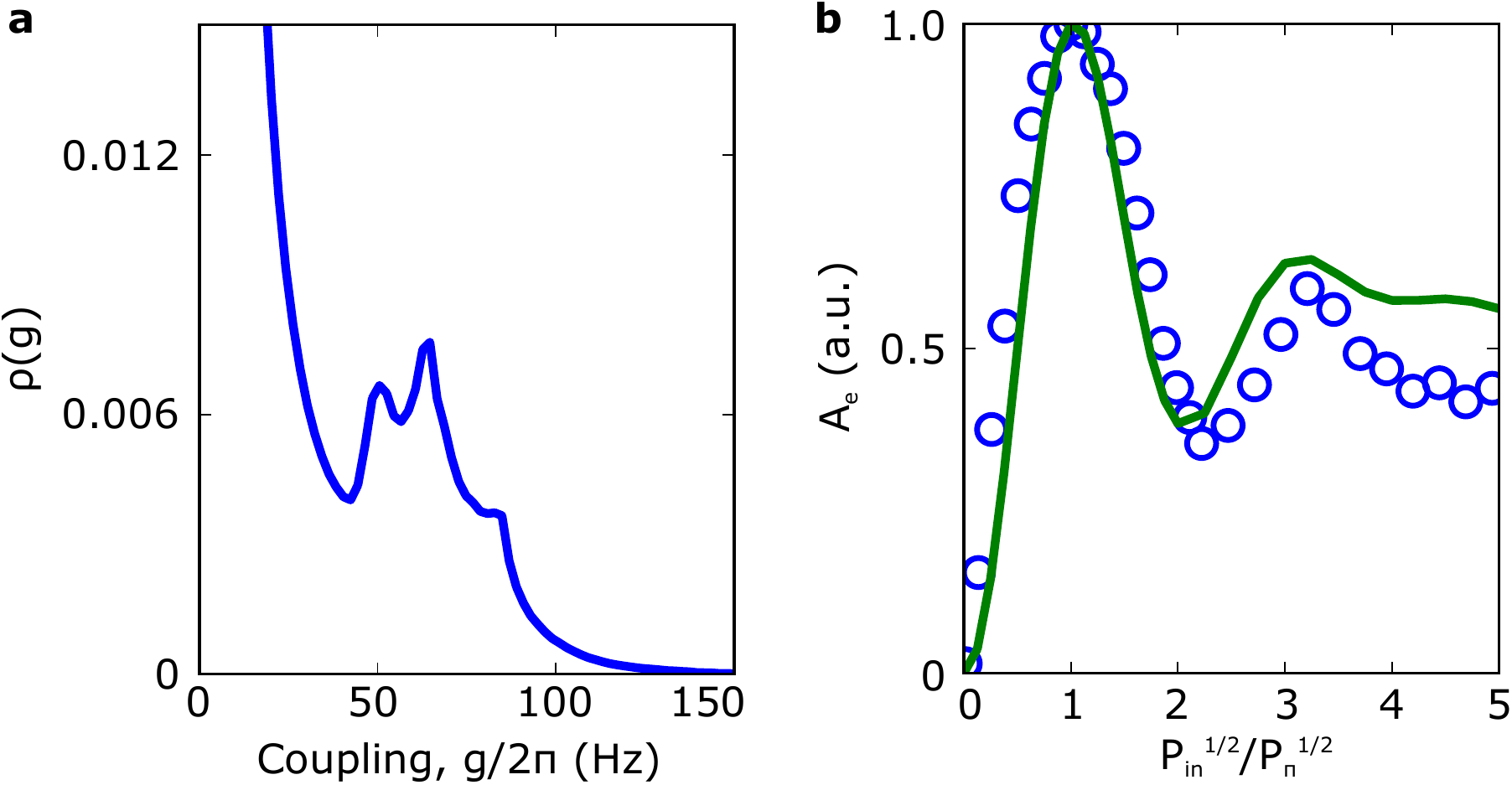}
  \caption{\label{fig3_ext}
  \textbf{Simulation of Rabi oscillations. a,} Distribution of the spin-cavity coupling g obtained from the spatial distribution of $\delta B_1$. \textbf{b,} Rabi oscillations measured at $B_0 =62.5$\,mT and $T =15$\,mK by varying the amplitude of the second pulse in the Hahn echo sequence (blue circles). The pulse amplitude is normalized to the value $P_\pi^{1/2}$ corresponding to the maximum in detected signal. The solid green line is the result of the numerical simulation of a spin ensemble described by $\rho(g)$.
  }
\end{figure}

\textbf{Estimate of single spin coupling g.} The interaction of the individual bismuth donor spins with the resonator mode of frequency $\omega_0$ is described by the Jaynes-Cummings Hamiltonian $H = \hbar g(\sigma^+a+\sigma^-a^+)$, where $\sigma^+$ ($\sigma^-$) is the spin raising (lowering) operator, $a$ ($a^+$) is the annihilation (creation) operator for intra-cavity photons, and $g$ is the spin-photon coupling strength. All the spin echo measurement in the present work, unless specified, are performed at 62.5\,mT on the $\ket{4,-1}\leftrightarrow\ket{5,0}$ and $\ket{4,0}\leftrightarrow\ket{5,-1}$ quasi-degenerate transitions whose matrix elements are equal to 0.28 and 0.22, respectively. As a consequence, two families of spins are probed simultaneously, with different coupling strengths given by $g = \gamma_e \bra{F,m}S_y\ket{F+1,m-1} \left\vert\delta B_1\right\vert$, where $\delta B_1$ is the magnitude of the magnetic field vacuum fluctuations at the spin location generated by the current vacuum  fluctuations $\delta i$ in the resonator inductor wire. The value of $\delta_i$ is given by $\delta_i = \omega_0\sqrt{\hbar/2Z_0}$, where $Z_0=\sqrt{L/C} =46\,\Omega$ is the characteristic impedance of the LC resonator estimated from electromagnetic simulation realized in CST Microwave studio. In order to obtain $\delta B_1$ in the silicon substrate around the inductor, we perform simulation of the magnetic field generated by $\delta i$ using COMSOL software. A non-uniform distribution of the current density in the superconducting niobium wire is assumed. The so obtained spatial distribution of the coupling constant $g(x,y)$ for the $\ket{4,0}\leftrightarrow\ket{5,-1}$ transition is shown in Fig.~\ref{fig1}b. Combining $g(x,y)$ for the two transitions with the spin implantation profile (see Fig.\,\ref{fig1}b), we extract the coupling distribution density of the spin ensemble $\rho(g)$ (see Fig.~\ref{fig3_ext}a) that we use in the spin simulation to reproduce the Hahn echo measurements.

\textbf{Spin simulation.} The simulation of the dynamics of the spin ensemble weakly coupled to the intra-cavity field consists in numerical solution of the system semi-classical equation of motion, as described in~\cite{ranjan_pulsed_2019}. The system of differential equations is solved for $M$ values of coupling $g$ and the results are averaged using $\rho(g)$ as weighting function. The transverse relaxation time is fixed to the measured $T_2$ while the longitudinal relaxation rate is imposed for each $g$ value to be the Purcell rate: $\Gamma_1 = \frac{\kappa g^2}{\kappa^2/4+\delta^2}$, where $\delta = \omega_s-\omega_0$ is the spin-cavity detuning. The spin distribution in frequency is modeled as a square centered at $\omega_0$ and 3\,MHz wide. Simulation of Rabi oscillations driven by varying the amplitude of the second pulse in Hahn echo sequence is shown in Fig.~\ref{fig3_ext}b. The average of signals emitted by spins with different Rabi frequencies results in damped oscillations that are in qualitative agreement with the experimental data. The pulse amplitude giving the maximum signal is then chosen as $\pi$ pulse amplitude for the simulation of the inversion recovery sequence.

\textbf{Pulse sequences.} All the $A_e$ data points presented in this work, excepted the SNR measurement, are obtained using an Hahn echo sequence followed by a Carr-Purcell-Meiboom-Gill sequence (CPMG). From 100 to 300 $\pi$ pulse are therefore applied after the first echo to record as many additional echoes, that are then averaged to improve the signal-to-noise ratio. The $\pi$ pulse length is fixed to 250\,ns and double with respect to the $\pi/2$ pulse. The amplitude of the two pulses is always the same while the phase of the $\pi$ pulses is $90^{\circ}$ shifted. The spectrum presented in Fig.~\ref{fig1}d is measured with a delay $\tau =8$\,\textmu s between the $\pi/2$ pulse and the $\pi$ pulse, 10 additional CPMG $\pi$ pulses separated by 16\,\textmu s and a repetition time $t_{rep}$ of 5.8\,s. The measurement is averaged 32 times at each $B_0$ field value. The $\Gamma_1$ measurements of Fig.~\ref{fig2} are realized with $\tau=15$\,\textmu s, 300 CPMG pulses separated by $2\tau$, $t_{rep}=80$\,s. At each temperature the $\Gamma_1^{-1}$ measurement is repeated 8 times and then averaged. The echo amplitude as a function of temperature shown in Fig.~\ref{fig2}c is obtained with $\tau =15$\,\textmu s, $t_{rep} =120$\,s, 300 CPMG pulses separated by $2\tau$ and no further averaging. For the $\Gamma_1$ measurement of Fig.\,\ref{fig3} $\tau = 15$\,\textmu s, $t_{rep} =120$\,s, 100 CPMG pulses are separated by $2\tau$ and no further averaging is used. In the $T_2$ experiment of Fig.~\ref{fig3}c, $t_{rep} = 10$\,s, 300 CPMG pulses separated by 8\,\textmu s are used and each data point is avaraged 4 times. The pulse sequence of each echo sample in the SNR experiment of Fig.~\ref{fig3}d does not include CPMG and has $\tau =15$\,\textmu s. Fig.~\ref{fig4}a shows results of $\Gamma_1$ measurements realized with $\tau =15$\,\textmu s, $t_{rep} =40$\,s, 300 CPMG pulses separated by $2\tau$. At each LED $I$ the $\Gamma_1$ measurement is repeated 4 times and then averaged. The cooling and recovery dynamics of Fig.~\ref{fig4}b are instead measured using $\tau =19.5$\,\textmu s, $t_{rep} =30$\,s, 100 CPMG pulses separated by $2\tau$ and 20 averages of each data point.

\end{document}